\documentclass[reprint, showpacs]{revtex4-1}
\usepackage{graphicx}
\usepackage{amsmath}
\usepackage{natbib}

\newcommand{\average}[1]{\left<#1\right>}
\newcommand{\skalarn}{\mathbf{r}'\cdot\mathbf{v}'}

\newcommand{\zunanji}{\mathbf{r}'\wedge\mathbf{v}'}

\newcommand{\hitr}{\mathbf{v}}

\newcommand{\lega}{\mathbf{r}}

\begin{document}
\title{Fermi acceleration in chaotic shape-preserving billiards}
\author{Benjamin Batisti\'c}

\affiliation{CAMTP - Center for Applied Mathematics and Theoretical
Physics, University of Maribor, Krekova 2, SI-2000 Maribor, Slovenia}

\date{\today}


\begin{abstract}
We study theoretically and numerically 
  the velocity dynamics of fully chaotic time-dependent shape-preserving
  billiards. The average velocity of an ensemble of initial conditions 
  generally asymptotically follows
  the power law $\average{v}=n^\beta$ with respect to the number of
  collisions $n$. If a shape of a fully chaotic time-dependent billiard is not
  preserved it is well known that the acceleration exponent is
  $\beta=1/2$. We show, on the other hand, that if a shape of a fully
  chaotic time-dependent billiard is preserved then
  there are only three possible values of $\beta$ depending solely on the
  rotational properties of the billiard. In a special case when
  the only transformation is a uniform rotation there is no acceleration,
  $\beta=0$. Excluding this special case, we show that 
  if a time-dependent transformation of a billiard is such that the 
  angular momentum of the billiard is
  preserved, then $\beta=1/6$ while $\beta=1/4$ otherwise. 
  Our theory is
  centered around the detailed study of the energy fluctuations in the adiabatic limit.
  We show that three quantities, two scalars and one tensor,
  completely determine the energy fluctuations of the billiard for arbitrary
  time-dependent shape-preserving transformations.
  Finally we provide several interesting numerical examples all in a
  perfect agreement with the theory.
\end{abstract}

\maketitle

\section{Introduction}

Because of their simplicity and generality, the
billiards are one of the most important dynamical systems. 
They are used as a model system in various
fields of research in classical and quantum mechanics. 
Billiards are especially convenient for numerical computation but they can be 
realized also experimentally, for example as a micro wave cavity,
acoustic resonators, optical laser resonators and quantum dots
\cite{Stoe}, which is in the domain of quantum chaos, but the
classical dynamics is important in the semiclassical picture. 

A time-dependent billiard was first considered as a model of a cosmic ray
particles 
acceleration process proposed by Fermi \cite{Fermi1949} and established by Ulam
\cite{Ulam1961}. 
It was argued at the time that the moving wall
accelerates the particle without limit. Such acceleration is called Fermi
acceleration. Now it is well known that Fermi acceleration does not
necessarily take place, for example in 1D system if the motion of the
walls is sufficiently smooth \cite{Lieberman1972}. However, Fermi acceleration
exists in almost all 2D billiards.
Numerical result show that the average velocity of an ensemble of
particles asymptotically
follows the power law $\average{v}=n^\beta$ where $\beta$ is the 
{\it acceleration exponent}.
In billiards with fully chaotic dynamics one would intuitively expect that
due to the loss of correlations between the successive velocity changes the
acceleration exponent equals $\beta=1/2$ in analogy with the random walk
process. This intuitive result is theoretically well supported
\cite{Gelfreich2008,Gelfreich2012}.
In addition it is shown that there exist trajectories 
with measure zero which accelerate even exponentially in continuous time ($\beta=1$). 
However, 
billiards can be transformed in such a special way that despite chaos,
$\beta$ can be smaller than $1/2$ and even zero,
as shown in this paper.
Various values of $\beta$ between 0 and 1 were found in systems with a coexisting
regular and chaotic motion \cite{Leonel2009a}. There is a strong numerical and theoretical
evidence that in such systems the exponential acceleration may become
predominant \cite{Shah2013,Gelfreich2013arxiv}. 
On the other hand the numerical studies of the time-dependent 
not shape-preserving elliptical billiard \cite{Lenz2009,Lenz2010,Oliveira2011a}, which is the 
integrable system as static, show that $\beta$
asymptotically equals $1/2$ while it passes a long transient regime where
acceleration is sub-diffusive i.e. $\beta<1/2$.

One of the basic assumptions of the theory of Gelfreich et al.
\cite{Gelfreich2008}, which predicts $\beta=1/2$ for fully chaotic time
dependent systems,
is the existence
of pairs of periodic orbits with a heteroclinic connection where their relative lengths undergo different
time evolutions. However, in shape-preserving time-dependent billiards
this assumption does not hold (at least not to the same order of
magnitude), and as a consequence $\beta<1/2$, as shown theoretically and
numerically in this paper.
A shape preserving transformation can be only a combination of rotation,
translation and scaling. The scaling transformations alone were already
studied in our previous work \cite{Batistic2011,Batistic2012}, but here we provide a complete theory for a
general shape preserving transformation.

In this paper we derive a differential equation for the velocity of a particle in a
reference frame in which a billiard is at rest.
This differential equation is used to derive a general formula
for the time evolution of energy fluctuations
on adiabatic time scales. We show that the order of magnitude of the energy fluctuations 
depend on whether a transformation is such that the angular momentum of
a billiard (assuming a constant mass) is preserved or not. 
We derive also the corresponding
asymptotic acceleration exponents.
We show that there is no acceleration ($\beta=0$) if
the only transformation of a billiard is a uniform rotation, which is a
counter example of the LRA conjecture \cite{Loskutov1999}, 
stated as follows:
"Thus, on the basis of our investigations we can advance the 
following conjecture: chaotic dynamics of a billiard with a 
fixed boundary is a sufficient condition for the Fermi acceleration
in the system when a boundary perturbation is introduced."
If this statement has to be understood as including the rigid/uniform
rotation, which is a specific perturbation of the billiard boundary,
then, in this sense, the result $\beta=0$ in any at least partially chaotic
billiard (when static) clearly violates the LRA conjecture.
Excluding this special example we show that the value of the acceleration
exponent is either $\beta=1/6$
if the angular momentum of the billiard is preserved or $\beta=1/4$
otherwise. 
Theoretical results are finally confirmed by the numerical
results.

\section{Generalities}

A billiard is a dynamical system in which a particle alternates between
the force-free motion and instant reflections from a
boundary. A boundary is a closed
curve in the configuration space
representing an infinite potential barrier.

Between collisions the particle velocity $\hitr$ is constant while
at collisions it changes instantly by a
\begin{equation}
  \Delta\mathbf{v}=R\left(\mathbf{v}-\mathbf{u}\right),
  \label{eqOdboj}
\end{equation}
where $\mathbf{u}$ is a velocity of a boundary at a collision point
and $R$ is a reflection tensor
\begin{equation}
  R=-2\,\mathbf{n}\,\mathbf{n}^{T},
  \label{eqTenzor}
\end{equation}
where $\mathbf{n}$ is a normal unit vector to the boundary at a collision point.
The reflection tensor is symmetric and satisfies
\begin{equation}
  R^{T}R = -2R.
  \label{eqTenzorProp}
\end{equation}
Using the properties of $R$ it is straightforward to show that
the norm $\|\hitr-\mathbf{u}\|$ is preserved at collisions.
Thus, in general, if $\mathbf{u}\neq0$, the norm of the velocity vector $v=\|\hitr\|$ is not preserved.
If $\mathbf{u}$ is constant and zero for every point on a boundary, then
the billiard is static. In a static billiard a magnitude of a particle
velocity is constant. 

\section{Primed space}

Consider a coordinate system $\lega'=\lega'(\lega,t)$ in which a billiard
boundary is at rest and name it
a primed space $S'$. The velocity of a point in $S'$ equals
\begin{equation}
  \hitr' = \frac{d\lega'}{dt} =
  \frac{\partial{\lega'}}{\partial{t}}+J\,\hitr,
  \label{eq:crtkana_hitrost}
\end{equation}
where 
\begin{equation}
  J = \left(\frac{\partial{\lega'}}{\partial{\lega}}\right)
  \label{eq:jakobi_crtkani}
\end{equation}
is the Jacobian of the coordinate transformation and $\hitr$ is the
velocity in the physical (untransformed) space.
By the definition of $S'$, the points on the billiard
boundary must satisfy $\hitr'=0$, thus having the velocity
\begin{equation}
  \mathbf{u} = -J^{-1}\,\frac{\partial{\lega'}}{\partial{t}},
  \label{eq:hitrost_biljarda}
\end{equation}
where $\lega'$ is a corresponding position of a boundary point in the
primed space.
At collisions the particle velocity $\hitr'$ changes by a
\begin{eqnarray}
  \Delta{\hitr'} &=& J\,\Delta{\hitr} \nonumber\\
                 &=& J\,R\,\left(\hitr-\mathbf{u}\right) \nonumber\\
                 &=& J\,R\,J^{-1}\,\hitr',
  \label{eqDeltaPrimedV}
\end{eqnarray}
where in the last equality we expressed $\hitr$ from
(\ref{eq:crtkana_hitrost}) and took into account
(\ref{eq:hitrost_biljarda}).

\section{Conformal transformations}

If $J$ is such that
\begin{equation}
  J\,R\,J^{-1} = R,
  \label{eqTenzorTransf}
\end{equation}
then according to (\ref{eqDeltaPrimedV}), $\hitr'$ obeys the
reflection law of a static billiard
\begin{equation}
  \Delta{\hitr'}=R\,\hitr',
  \label{eqOdbojniZakon}
\end{equation}
which implies that the norm $v'=\|\hitr'\|$ is preserved at collisions.

Using (\ref{eqTenzorTransf}) in (\ref{eqTenzorProp}) gives the relation
\begin{equation}
  \left(J^{T}J\right)^{-1}R^{T}\left(J^{T}J\right)R=-2\,R,
  \label{eqTransfRrelacija}
\end{equation}
which is satisfied only if
\begin{equation}
  J^TJ = \alpha\,I,
  \label{eqtransfRpogoj}
\end{equation}
where $I$ is the identity matrix and $\alpha$ is a constant. We can determine
$\alpha$ by taking
the determinant of both hand sides of (\ref{eqtransfRpogoj}).
In 2D these gives $\alpha^2=\left|J\right|^2$ and the relation
\begin{equation}
   J^TJ=\left|J\right|.
  \label{eqKonfPogoj}
\end{equation}
Transformations satisfying (\ref{eqKonfPogoj}) are the angle-preserving 
or conformal transformations
which are everywhere a combination of a scalar multiplication and
a rotation.

\section{Shape-preserving transformations}

In this paper we study only shape-preserving transformations which are
linear conformal transformations.
By definition, shape-preserving transformations 
preserve curvatures and angles. The Jacobian of a shape-preserving
transformation is independent of position.  
A most general shape-preserving transformation has the form 
\begin{equation}
  \lega'=J\left(\lega-\mathbf{z}\right),
  \label{eqGtrans}
\end{equation}
where $\mathbf{z}$ is a translation vector,
and $J$ is the Jacobian of the form
\begin{equation}
  J=\left(q\,O\right)^{-1},
  \label{eqJacShapeP}
\end{equation}
where $q$ is a scaling factor and 
\begin{equation}
O=
\left(
\begin{array}{cc}
\cos{\phi} & -\sin{\phi} \\
\sin{\phi} & \cos{\phi}
\end{array}
\right)
\label{eqRotM}
\end{equation}
is a rotation matrix.

Trajectories in $S'$ are curved if a shape-preserving
transformation depends on time.
However, because
shape-preserving transformations transform
straight lines into straight lines,
the curvature must vanish in the
adiabatic limit.
In this limit the curvature of the trajectories plays no role anymore,
but the magnitude of the velocity is still governed by the transformation.
A curvature radius of a trajectory in $S'$ equals
\begin{equation}
  \mathcal{R} = \frac{ds'}{d\phi'}=
  \frac{v'\,dt}{\|\hitr'\times\dot{\hitr}'\,dt\|/v'^2}
  =\frac{v'^3}{\|\hitr'\times\dot{\hitr}'\|},
  \label{eqCurvature}
\end{equation}
where 
\begin{equation}
  \dot{\hitr}'=2\,\dot{J}\,J^{-1}\,\hitr'
  -2\,\dot{J}\,J^{-1}\,\frac{\partial\,\lega'}{\partial{t}}
  +\frac{\partial^2\,\lega'}{\partial{t}^2}
  \label{eqPrimedVdot}
\end{equation}
is the acceleration in $S'$.
The dot denotes the time derivative throughout this paper.
In the adiabatic limit, when $v'\rightarrow{\infty}$, 
$\mathcal{R}$ is in general proportional to $v'$, except when
$\dot{J}\,J^{-1}$ is diagonal (no rotations) 
and $\mathcal{R}$ is proportional to $v'^2$.
In any case $\mathcal{R}$ diverges in the adiabatic limit.

Vanishing curvatures and the law of reflection (\ref{eqOdbojniZakon}) lead to the important
conclusion that in the adiabatic limit
the geometry of trajectories in $S'$ of a
shape-preserving time-dependent billiard approaches the velocity independent geometry
of trajectories of the corresponding static billiard.
This fact allows us to investigate the dynamics of a shape-preserving
billiard much deeper then in a general time-dependent billiard. We can take the trajectories of a 
static billiard as an approximation for the trajectories in $S'$ thus
reducing the system of four differential equations to a single differential
equation for $v'$.

We construct the differential equation for $v'$ from $v'\,\dot{v}'=\hitr'\cdot\dot{\hitr}'$ using
(\ref{eqGtrans}), (\ref{eqJacShapeP}) and (\ref{eqPrimedVdot}),
\begin{equation}
  v'\,\dot{v}'= -\frac{2\,\dot{q}}{q}\,v'^2+
  \left(\omega^2+\frac{\ddot{q}}{q}\right)\,\skalarn
  + \frac{\dot{\Gamma}}{q^2}\,\zunanji
-J\,\ddot{\mathbf{z}}\cdot\hitr'
\label{eqCentVdiffEq}
\end{equation}
where $\omega=\dot{\phi}$ is the angular velocity of a billiard, 
\begin{equation}
  \Gamma = \omega\,q^2
  \label{eqBilVrtKol}
\end{equation}
is the quantity proportional to the angular momentum of the billiard 
and
\begin{equation}
 \zunanji\equiv r'_x\,v'_y-r'_y\,v'_x\equiv v'\,r'\,\sin{\alpha'} 
  \label{eqZunanji}
\end{equation}
 is the angular momentum of the particle in $S'$.

We multiply (\ref{eqCentVdiffEq}) by $q^4$, move the first term from the
right hand side to the left hand side and write the left hand side as a
total derivative
\begin{equation}
  \frac{d}{dt}\left(\frac{q^4\,v'^2}{2}\right) = B\,\zunanji + A\,\skalarn -
  \mathbf{a}\cdot{\hitr'},
  \label{eqVdiffEqB}
\end{equation}
where we introduced
\begin{eqnarray}
  \label{eqBdef}
  B&=& q^2\,\dot{\Gamma},\\
  \label{eqAdef}
  A&=& \Gamma^2-q^3\,\ddot{q},\\
  \label{eqAccDef}
  \mathbf{a}&=& q^3\,O^{-1}\,\ddot{\mathbf{z}}.
\end{eqnarray}

The fact that $\skalarn$ and $\hitr'$ are total time-derivatives 
of $r'^2/2$ and $\lega'$ respectively allows us to rewrite (\ref{eqVdiffEqB}) as
\begin{equation}
  \frac{d}{dt}\left(
  \frac{q^4\,v'^2}{2}-\frac{A\,r'^2}{2}+\mathbf{a}\cdot\lega' \right) 
  = B\,\zunanji - \frac{\dot{A}\,r'^2}{2} +
  \dot{\mathbf{a}}\cdot{\lega'}.
  \label{eqVdiffEqA}
\end{equation}
Note that $\zunanji$ is not a total time-derivative. 

In the adiabatic regime of sufficiently large $v'$ we can consider
$q^4\,v'^2/2$ as the dominant term in (\ref{eqVdiffEqA}), thus neglecting
all the other terms we conclude that 
\begin{equation}
q^4\,v'^2 = \textrm{constant},
\label{eqAdConst}
\end{equation}
which is actually a more accurate version of the well known adiabatic
theorem for ergodic time-dependent billiards
\cite{Hertz1910,Einstein1911},
\begin{equation}
  v\,\sqrt{\mathcal{A}} = \textrm{constant}.
    \label{eqInvarianta}
\end{equation}
where $\mathcal{A}$ is the billiard area. Equation (\ref{eqInvarianta}) 
follows from (\ref{eqAdConst}) after the approximation $v'\approx v/q$ and substitution
$q^2\propto\mathcal{A}$. It is important to note that the adiabatic
invariant (\ref{eqInvarianta}) is valid for all shape-preserving
billiards, regardless of whether they are ergodic or not.
The adiabatic invariant describes the evolution of the average velocity of
an ensemble on the adiabatic time scales. The width of the velocity distribution
is spreading around its average and eventually results in the Fermi
acceleration.
This paper provides an accurate theoretical description of this
process. 

If $B=\dot{A}=0$ and $\dot{\mathbf{a}}=0$ then (\ref{eqVdiffEqA}) can
be integrated exactly.
It follows from (\ref{eqBdef}) that
$B=0$ if a transformation is such that the angular momentum of
a billiard is preserved, which is when $\Gamma=\omega\,q^2=\text{constant}$.
Using this in (\ref{eqAdef}) we see that $\dot{A}=0$ when $B=0$ if
$q^3\,\ddot{q}=\text{constant}$, which is when $q$ is of the from 
\begin{equation}
  q=\sqrt{c_1+c_2\left(t+c_3\right)^2},
  \label{eqAconstCond}
\end{equation}
where $c_1$, $c_2$ and $c_3$ are constants.
Very interestingly though, if $q$ is of the form
(\ref{eqAconstCond}) and $\omega\propto 1/q^2$ and $\mathbf{a}=0$, then (\ref{eqVdiffEqA})
can be integrated exactly, but this driving is not periodic.

The only periodic and infinitely smooth solution of the condition
$B=\dot{A}=0$ and $\dot{\mathbf{a}}=0$ is a uniform rotation,
i.e. $\dot{\omega}=\dot{q}=0$ and $\mathbf{a}=0$. 
In this case 
\begin{equation}
v'^2-\omega^2\,r'^2=\text{constant}, 
\label{eqConLawRotat}
\end{equation}
as follows from
(\ref{eqVdiffEqA}). Thus $v'$ is bounded if $r'$ is bounded. 
Obviously, if $v'$ is bounded then $v$ is bounded as well, so
there is no Fermi acceleration in uniformly rotating billiards.
Thus we conclude that the acceleration exponent $\beta=0$.

Before we proceed we have to define several different dynamical time-scales relevant for
shape-preserving time-dependent billiards. 
In a static billiard the two relevant scales are the 
ergodic time-scale $\tau_E=l_E/v$ on which the particle uniformly visits the
whole accessible phase space and the time averages can be replaced by the
phase-space averages, and the correlation time scale $\tau_C=l_C/v$
on which the autocorrelations vanish. The characteristic geometrical
lengths $l_E$ and $l_C$ are independent of the particle
velocity and the corresponding time-scales are vanishing in the limit
$v\rightarrow\infty$. The next relevant scale is the adiabatic time-scale $\tau_A$,
defined as a scale on which the variations of the 
adiabatic invariant (\ref{eqAdConst}) remain relatively small. Adiabatic time-scale $\tau_A$ is 
an increasing function
of $v'$ and can be made arbitrarily large. 
We shall denote with $\tau_B$ a time-scale of a billiard motion,
proportional to $1/u_{\max}$ where $u_{\max}$ is a maximal velocity of the
boundary. 

Suppose the observation time $t$ is much smaller than $\tau_A$ and
$\tau_E\ll \tau_B$. 
We introduce
three quantities which 
have a zero mean by construction:
\begin{eqnarray}
  \varsigma &=&  r'\sin{\alpha'} - \average{r'\sin\alpha'},\\
  \eta &=&  r'^2/2-\average{r'^2/2},\\
  \xi &=&  \lega'-\average{\lega'}.
  \label{eqDefVariables}
\end{eqnarray}
Where $\langle\,\rangle$ denotes a phase-space average.
Using these quantities we write (\ref{eqVdiffEqA}) in the form
\begin{equation}
  E-E_0=\int_0^t\,dt\,P,
  \label{eqEintegEq}
\end{equation}
where 
\begin{align}
  E&=\frac{q^4\,v'^2}{2}-q^4\,\omega\,\,v'_A\average{r'\sin{\alpha'}}\nonumber\\
   &\quad-\frac{A}{2}\,\left(r'^2-\average{r'^2}\right) + 
  \mathbf{a}\cdot\left(\lega'-\average{\lega'}\right)
  \label{eqCentEnerDef}
\end{align}
is interpreted as the energy, $E_0$ is its initial value, and 
\begin{equation}
  P=B\,v'_A\,\varsigma-\dot{A}\,\eta+\dot{\mathbf{a}}\cdot\xi
  \label{eqPowDef}
\end{equation}
is interpreted as the power.
Here we have made the approximation by substituting $v'$ in the low order term 
$B\,\zunanji=B\,v'\,r'\sin\alpha'$, with the adiabatic approximation
\begin{equation}
  v'_A=\frac{q_0^2\,v'_0}{q^2},
  \label{eqVadiabaticDef}
\end{equation}
which comes from (\ref{eqAdConst}).

We are interested in the statistical properties of the energy fluctuations
$\delta{E}=E-E_0$, in particular in the
second moment $\average{\delta{E}^2}$. 
We assume that the quantities $\varsigma$, $\eta$ and $\xi$ are mutually
uncorrelated and that in a regime where $\tau_C\ll
\tau_B$ their autocorrelation functions can be approximated
with the Dirac delta distributions,
\begin{eqnarray}
  \label{eqBcorr1}
  \average{\varsigma(t_1)\,\varsigma(t_2)}&=&
  \frac{\kappa_\varsigma}{v'}\,\delta(t_2-t_1),\\
\label{eqBcorr2}
  \average{\eta(t_1)\,\eta(t_2)}&=&
  \frac{\kappa_\eta}{v'}\,\delta(t_2-t_1),\\
  \label{eqBcorr3}
  \average{\mathbf{\xi}(t_1)\,\mathbf{\xi}^T(t_2)}&=& 
  \frac{K}{v'}\,\delta{(t_2-t_1)},
\end{eqnarray}
where we have introduced numbers $\kappa_\varsigma$ and $\kappa_\eta$ and
a tensor $K$. We expect that in the adiabatic limit
$\kappa_\varsigma$, $\kappa_\eta$ and
$K$ are the same as in the static billiard and thus independent of the
velocity. Note that these three quantities are also independent of the
driving.

Taking into account the autocorrelation functions we find
the following formula for the time evolution of the second moment of the
energy fluctuations,
\begin{eqnarray}
  \average{\delta{E}^2}&=& \average{\int_0^t\int_0^t\,dt_1\,dt_2\,
    P(t_1)\,P(t_2)}\nonumber\\
  &=&
  \int_0^t\,\frac{dt}{v'_A}\,\left(v'^2_A\,B^2\,\kappa_\varsigma+\dot{A}^2\,\kappa_\eta
  +\dot{\mathbf{a}}\,K\,\dot{\mathbf{a}}\right).
  \label{eqCentEdiff}
\end{eqnarray}
Again, here we made the approximation $v'\approx v'_A$.
Since all the terms under the  integral are non-negative, $\langle
\delta{E}^2\rangle$ is a strictly increasing function of time except when
the billiard boundary is at rest and the integrand is zero.

We distinguish cases when the angular momentum of the billiard
is preserved ($B = 0$) and when it is not ($B\neq0$).
If $B\neq0$, then after neglecting small terms $\dot{A}^2\,\kappa_\eta$
and $\dot{\mathbf{a}}\,K\,\dot{\mathbf{a}}$,
\begin{equation}
  \average{\delta{E}^2}\approx
  \int_0^t\,dt\,v'_A\,B^2\,\kappa_\varsigma.
  \label{eqEvarB}
\end{equation}
On the other hand, if $B=0$, (\ref{eqEvarB}) vanishes and we have to deal
with the previously neglected terms only
\begin{equation}
  \average{\delta{E}^2}=
  \int_0^t\,\frac{dt}{v'_A}\,\left(\dot{A}^2\,\kappa_\eta
  +\dot{\mathbf{a}}\,K\,\dot{\mathbf{a}}\right).
  \label{eqEvarA}
\end{equation}
The basic difference between the two cases is that with the increasing
$v'$ the variance of the energy fluctuations $\average{\delta{E^2}}$ 
grows faster with time if $B\neq0$ and slower if $B=0$. In a case where 
$B=0$, $\dot{A}=0$ and $\dot{\mathbf{a}}=0$ we
see that $\average{\delta{E}^2}$ is constant. For a uniform rotation with
the conserved angular momentum ($B=0$) this again implies no Fermi
acceleration and $\beta=0$. Note that once the
quantities $\kappa_\varsigma$, $\kappa_\mu$ and $K$ are determined for a
billiard, they can be used to derive $\average{\delta{E}^2}$ for arbitrary
drivings.

\begin{figure}
  \centering
  \includegraphics[scale=1]{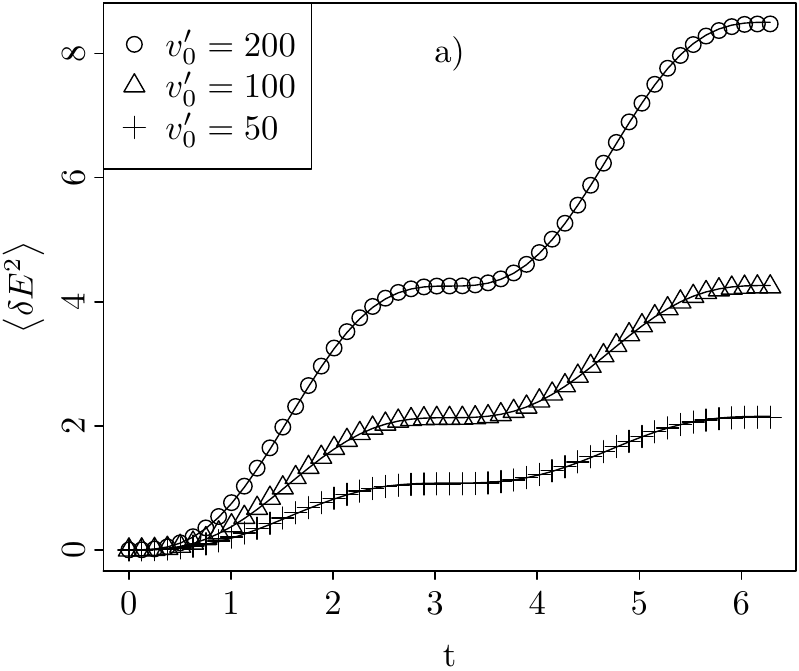}
  \hfill
  \includegraphics[scale=1]{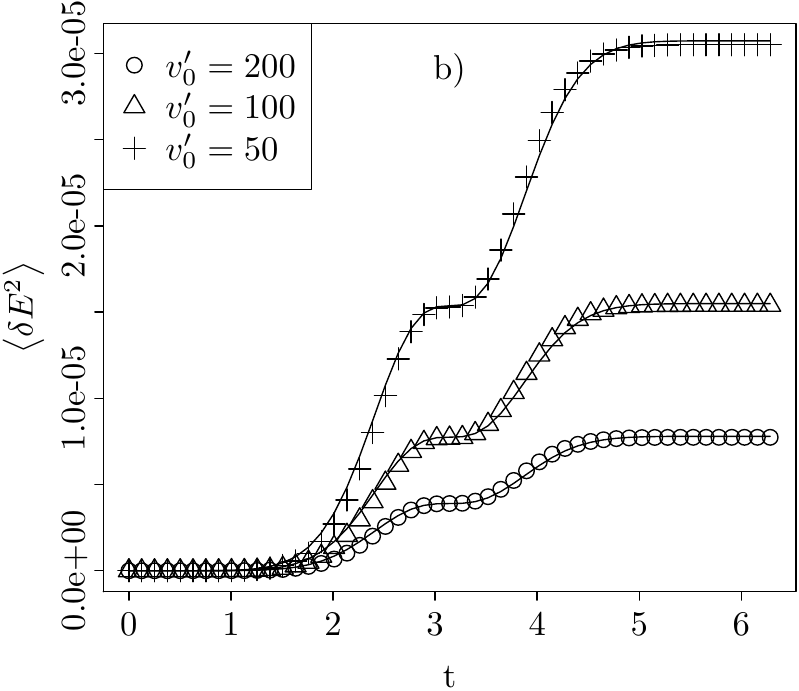}
  \caption{A time evolution of
    $\average{\delta{E^2}}$ for several different $v'_0$ and two different
    drivings of the Sinai billiard.
    \textbf{a)} The billiard rotates with $\omega=\cos{t}$. 
    \textbf{b)} The billiard rotates with $\omega=1+0.2\,\cos{t}$,
    but it is also scaled by a factor $q$
    in such a way that the angular momentum of the billiard is constant: 
    $\omega\,q^2=1$  and $B=0$. Symbols represent numerical results while
  lines are the theoretical predictions. The agreement is very good. Note
that we show the time evolution of $\average{\delta{E}^2}$ only within the
first period of driving
and that $\average{\delta{E}^2}$ continues to increase in the same
oscillatory manner, just as predicted by the theory (\ref{eqCentEdiff}).}
  \label{figRotacija}
\end{figure}

\begin{figure*}
  \includegraphics[scale=1]{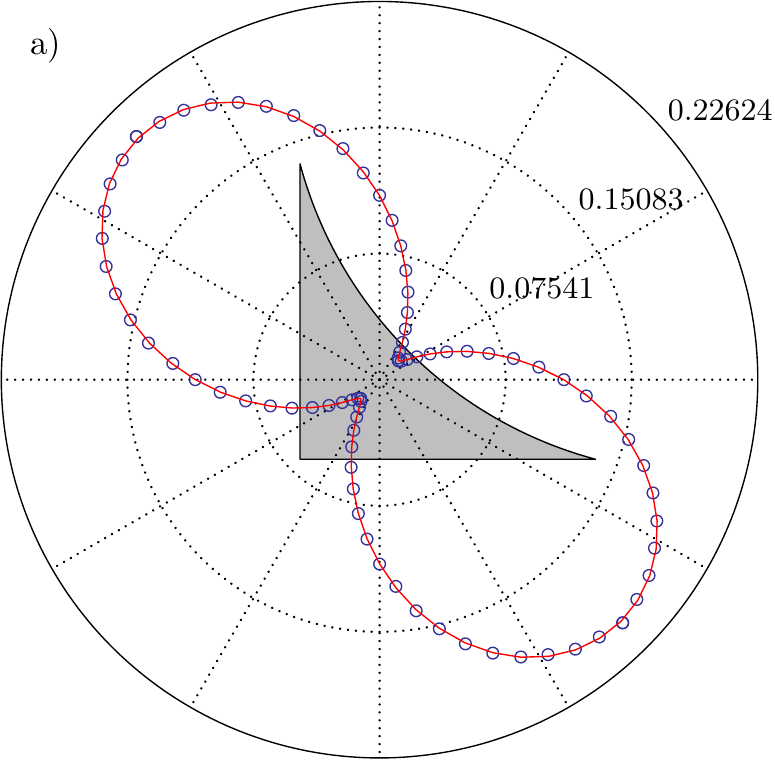}
  \hfill
  \includegraphics[scale=1]{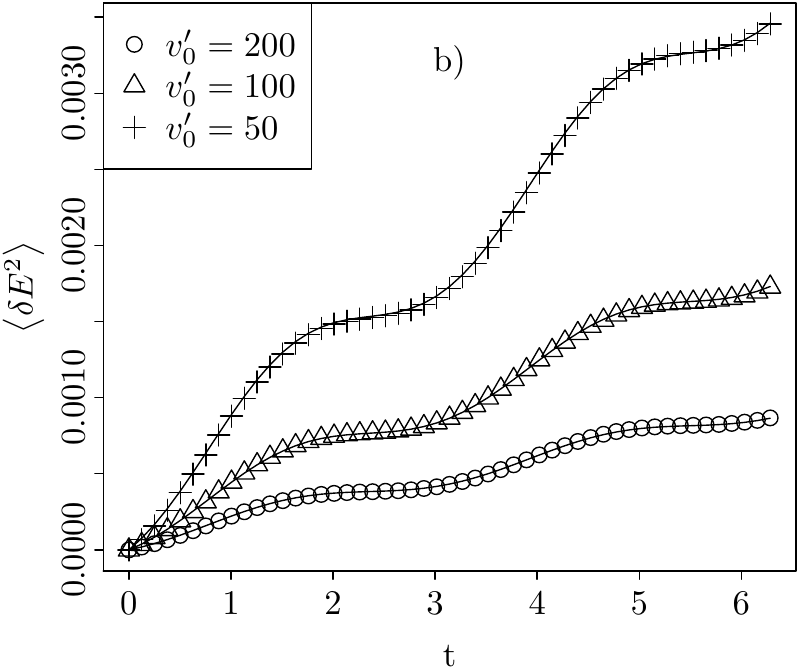}
  \caption{(Color online) A translating Sinai billiard. \textbf{a)} A tensor $K$ in a
  polar representation $\kappa(\phi)=a+b\,\sin{2\,\phi}$, with $a=0.11014$ and
$b=-0.09416$ (red line through circles), as determined numerically (circles) by shaking the
billiard in different directions $\phi$ like
$\mathbf{z}=0.2\,\cos{t}\left(\cos{\phi},\,\sin{\phi}\right)$ with
$v'_0=100$. For convenience, the billiard shape and its orientation with
respect to the tensor is represented with the shaded area. 
\textbf{b)} A time evolution of $\average{\delta{E^2}}$
for three different $v'_0$ in a circularly translating Sinai billiard
where the translation vector equals $\mathbf{z}=0.5\left(\cos{t},\,\sin{t}\right)$. 
The solid lines are theoretical curves (no fitting)
$\average{\delta{E^2}}=0.25\left(a\,t-b\,\sin^2{t}\right)/v'_0$.}
\end{figure*}

\section{Fermi acceleration}

When we follow the particle velocity on the long run, we observe that the
average velocity follows the adiabatic law (\ref{eqInvarianta}) or
(\ref{eqVadiabaticDef}), but in addition we see
diffusion in the velocity space, which eventually results in the Fermi
acceleration.

The evolution of the average velocity $\average{v}$ with respect to the
number of collisions $n$ of an ensemble of initial
conditions asymptotically follows the
power law
\begin{equation}
    \average{v}\propto n^\beta,
    \label{eqPowLaw}
\end{equation}
where $\beta$ is the acceleration exponent.
This law is empirically well established
\cite{Carvalho2006b,Leonel2009a,Kamphorst2007,Oliveira2012a}.
The Fermi acceleration is directly linked to the velocity diffusion
process. As we shall see, $\beta$ is determined by the way how the
diffusion constant depends on the velocity, describing the inhomogeneous
diffusion in the velocity space.
The acceleration exponent $\beta$ can be deduced 
from the time evolution of a second moment of the velocity
fluctuations $\delta{v'}=v'-v'_A$. 
From $\delta{E}\approx\delta{T}$ we have
\begin{equation}
  \delta{E}\approx\frac{q^4\,\left(q_0^2\,v'_0/q^2+\delta{v'}\right)^2}{2}
  -\frac{q_0^4\,v'^2_0}{2}\approx
  \left(q\,q_0\right)^2\,v'_0\,\delta{v'},
  \label{eqCentdEdV}
\end{equation}
from which it follows
\begin{equation}
  \average{\delta{v'}^2}=\frac{\average{\delta{E}^2}}{\left(q\,q_0\right)^4
  v'^2_0}.
  \label{eqCentVvar}
\end{equation}

We see from (\ref{eqCentVvar}) and (\ref{eqCentEdiff}) that the time
average of $\average{\delta{v'^2}}$ on the intervals $\tau$ satisfying
$\tau_B\ll \tau \ll t \ll \tau_A$, is a linearly increasing function of
time $t$, since the integrands may be considered as approximately constant,
which must be true for $\average{\delta{v}^2}$ as well and must be of the
form
\begin{equation}
  \average{\delta{v^2}}_\tau = \frac{2\,D\,t}{v^\mu},
  \label{eqCent2momentVt}
\end{equation}
where $D$ is some velocity independent constant and if the angular
momentum of the billiard is preserved ($B= 0$) then
according to (\ref{eqEvarA}) $\mu=3$ and if it is not ($B\neq0$) then according to
(\ref{eqEvarB}) $\mu=1$.

Now according to (\ref{eqCent2momentVt}) the evolution of the velocity
distribution $P(v)$ is described by the inhomogeneous diffusion equation
\begin{equation}
\frac{\partial}{\partial{t}}P(v)=D\,\frac{\partial}{\partial{v}}
\left(v^{-\mu}\,\frac{\partial}{\partial{v}}P(v)\right).
\label{eqDiffEq}
\end{equation}
We assume that after a long enough
time the shape of the velocity distribution $P(v)$ is independent of time.
 Let the shape of the velocity distribution
$P(v)$ be the same as the shape of some function $F(x)$ with the first
two moments equal to unity: $\int F(x)\,dx=1$ and $\int x\,F(x)\,dx=1$.
In this case we can write
\begin{equation}
P(v)=u^{-1}F\left(v\,u^{-1}\right),
\label{eqPforma}
\end{equation}
where the average velocity $u\equiv\average{v}$ is a function of
time. Putting (\ref{eqPforma}) in (\ref{eqDiffEq}) gives
\begin{equation}\label{eq13}
-\frac{\dot{u}}{u^{2}}\,\frac{\partial}{\partial{x}}\bigg(x\,F(x)\bigg)=
\frac{D}{u^{\mu+3}}\,\frac{\partial}{\partial{x}}\left(x^{-\mu}\,\frac{\partial}{\partial{x}}F(x)\right),
\end{equation}
where $x=v\,u^{-1}$. Because $F(x)$ does not depend on time, $u$ must
satisfy the following differential equation
\begin{equation}\label{eq14}
\dot{u}\,u^{\mu+1}=k\,D,
\end{equation}
where $k$ is a positive constant, which is found together with
$F(x)$ by solving the differential equation
\begin{equation}
  \frac{\partial}{\partial{x}}F(x)+k\,x^{\mu+1}\,F(x)=0
\label{eqDiffZaF}
\end{equation}
and imposing the normalization conditions. We find
\begin{equation}
  F(x) \propto e^{-k\,x^{\mu+2}}
  \label{eqFdistrib}
\end{equation}
and
\begin{equation}
  \average{v}=u \propto \left(D\,t \right)^{\frac{1}{\mu+2}}.
  \label{eqAccLaw}
\end{equation}
For the evolution of the mean velocity $\average{v}$ with respect to the number of
collisions $n$ it follows from
$\Delta{n}\propto\average{v}\Delta{t}$
after a straightforward manipulation
\begin{equation}
    \average{v}\propto n^{\frac{1}{\mu+3}}
    \label{eqAccLawN}
\end{equation}
and thus
\begin{equation}
    \beta = \frac{1}{\mu+3}.
    \label{eqBetaVsMu}
\end{equation}

From (\ref{eqCentVvar}) combined with (\ref{eqCentEdiff}) we see that the
diffusion exponent $\mu$ equals $1$ if $B\neq 0$ or $3$ if $B=0$ and the
corresponding acceleration exponents are $\beta=1/4$ and $\beta=1/6$.
Together with the $\beta=0$ for uniformly rotating billiards discussed
before, these exhaust
all possible values for $\beta$ in time-dependent shape-preserving
fully chaotic billiards. This is the central result of this paper.

\section{Numerical results}

For numerical computation we choose a completely chaotic Sinai billiard, defined as
an area between the coordinate axes and the circle $(x-a)^2+(y-a)^2=4$
where $a=\sqrt{2+\sqrt{3}}$, shown as a shaded area in figure 2a.

We shall consider four different drivings and test the validity of
(\ref{eqCentEdiff}), which describes the time evolution of the second
moment of the energy fluctuations.

The first example of the driving is a nonuniform rotation where the angle
of the billiard $\phi$ changes with time as $\phi=\sin{t}$.
According to
(\ref{eqGtrans}) in this case $B=\dot{\omega}=-\sin{t}$,
$A=\cos^2{t}$ and $\mathbf{a}=0$. For large velocities if $B\neq0$ we can
neglect terms involving $A$ and the equation (\ref{eqEvarB}) gives
for the time evolution of the variance of energy fluctuations
\begin{equation}
    \average{\delta{E^2}}=\kappa_\varsigma\,v'_0\,\int_0^t dt\,\sin^2{t}=
    \kappa_\varsigma\,v'_0\,\left(\frac{t}{2}-\frac{\sin{2\,t}}{4}
    \right).
    \label{eqRotNum}
\end{equation}
We took $10^6$ initial conditions at three
different initial velocities 
$v'_0=50,100,200$, at initial time $t_0=0$ and uniformly distributed on the
remaining 3D phase space.
Theoretical prediction (\ref{eqRotNum}) agrees with numerical results very well as
shown in figure 1a. In all cases $\kappa_\varsigma$ is approximately the
same $\kappa_\varsigma\approx0.0135$, as determined by the fitting
procedure, thus it is indeed independent of the velocity.
At large times we observe Fermi acceleration with $\beta=1/4$ as discussed
below.

The second example of the driving is a composition of scaling and rotation in such a way
that that the angular momentum of the billiard is preserved ($B=0$). 
The angular velocity is $\omega=1+0.2\,\cos{t}$ and
$q=1/\sqrt{\omega}$, such that $\Gamma=\omega\,q^2=1$ and
$B=q^2\,\dot{\Gamma}=0$. In this case we use the equation
(\ref{eqEvarA}). From (\ref{eqGtrans}) we find
\begin{equation}
    \dot{A}=-\frac{25\,\left(77\,\sin{t}-40\,\sin{2\,t}+\sin{3\,t}\right)}
    {8\,\left(5+\sin{t}\right)^5},
    \label{eqRotBreNum}
\end{equation}
which is used in (\ref{eqEvarA}) leading to
\begin{equation}
  \average{\delta{E}^2}=
  \frac{\kappa_\eta}{v'_0}\,\int_0^t dt\,\frac{q_0^2\,\dot{A}^2}{q^2}.
    \label{eqEvarAA}
\end{equation}
The analytical expression of this integral is too complicated to be
shown here, and it was evaluated numerically in practice as well. 
We took $10^6$ initial conditions at three
different initial velocities 
$v'_0=50,100,200$, at initial time $t_0=0$ and uniformly distributed on the
remaining 3D phase space.
A very good agreement with the theory is found as shown in figure 1b.
In all cases $\kappa_\eta$ is approximately the
same $\kappa_\eta\approx0.0183$, as determined by the fitting procedure, 
thus it is indeed independent of the
velocity.
At large times we observe Fermi acceleration with $\beta=1/6$ as discussed
below.

We used the third driving to find the tensor $K$ defined in
(\ref{eqBcorr3}). We translate the billiard back and forth 
as $\mathbf{z}=\left( \cos{\phi}, \sin{\phi}\right)\,\cos{t}$
in 20 different directions $\phi$ spanning from $-\pi/4$ to $\pi/4$.
According to (\ref{eqEvarA}) we have
\begin{equation}
  \average{\delta{E}^2}=
  \frac{\kappa(\phi)}{v'_0}\,\int_0^t dt\,\sin^2{t}=
  \frac{\kappa(\phi)}{v'_0}\,\left(\frac{t}{2}-\frac{\sin{2\,t}}{4}
  \right),
    \label{eqTransNum}
\end{equation}
where 
\begin{equation}
    \kappa(\phi)=(\cos\phi,\sin\phi)^T\,K\,(\cos\phi,\sin\phi).
    \label{eqDefKkapa}
\end{equation}
An ensemble of $10^6$ initial conditions at
$t_0=0$ and $v'_0=100$, uniformly distributed on the remaining phase space,
evolved one
period in time, was used to determine the evolution of
$\average{\delta{E^2}}$. Measurements of $\kappa(\phi)$ are shown as small
circles in a polar plot in figure 2a. 
Because of the symmetries of the billiard, tensor $K$ must be of
the form
\begin{equation}
    K=
\left( \begin{array}{cc}
a & b\\
b & a
\end{array} \right)
    \label{eqTenzroRep}
\end{equation}
and
\begin{equation}
    \kappa(\phi)=a+b\,\sin{2\phi}.
    \label{eqKapMal}
\end{equation}
Values of $a$ and $b$ were found by the best fit procedure. In figure 2a
we see that this model describes data very well.

The last example of the driving is a circular translation of the center of
mass of the billiard
where $\mathbf{z}=0.5\,(\cos{t},\sin{t})$, $q=1$ and $\omega=0$. Although the centre of mass of the
billiard is rotating around the origin of the coordinate system, the
billiard plane is not rotating, so the angular momentum of the billiard 
is zero and thus $B=0$ and we expect a slower diffusion.
From (\ref{eqEvarA}) and (\ref{eqTenzroRep}) we have
\begin{equation}
    \average{\delta{E^2}}=\frac{\left(a\,t-b\,\sin^2{t}\right)}{4\,v'_0}.
    \label{eqLast}
\end{equation}
We took $10^6$ initial conditions at three
different initial velocities 
$v'_0=50,100,200$, at initial time $t_0=0$ and uniformly distributed on the
remaining 3D phase space.
Now without any fit, using the values of $a$ and $b$ computed previously, 
we find a very good agreement with the numerical results as shown in
figure 2b.

\begin{figure}
  \centering
  \includegraphics[scale=1]{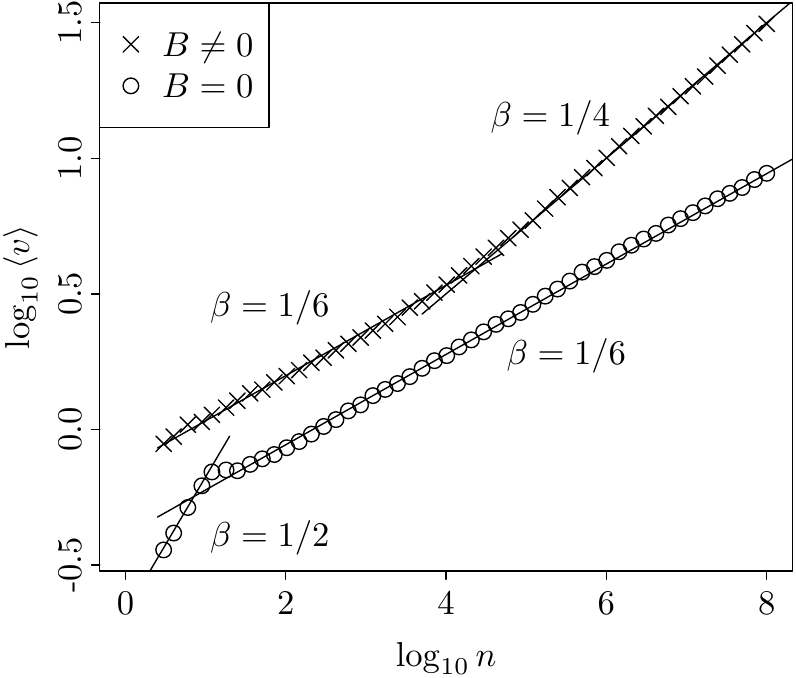}
  \caption{The mean velocity versus the number of collisions
  and the acceleration exponents for two different drivings: rotation
  (crosses) with $\phi(t)=\sin{t}$ and translations (circles) with
  $\mathbf{z}=(1/\sqrt{2},-1/\sqrt{2})\,\sin{t}$. In both we took ensemble
of $10^3$ initial conditions at $v_0=0.1$, $t_0=0$ and uniformly
distributed on the remaining phase-space. For very small velocities we
expect $\beta=1/2$. This regime is clearly visible
in translating billiard (circles). 
At bigger velocities $\beta=1/6$ in both cases. However, because
in the case of rotation $B\neq 0$ the asymptomatic value of $\beta$
eventually becomes $1/4$. The slopes of the lines have exact theoretical
values without fitting.}
  \label{figAcc}
\end{figure}

Finally we calculated the acceleration exponents. We took two different
drivings. One is a nonuniform rotation with $\phi(t)=\sin{t}$ and the other is
a translation with $\mathbf{z}=(1/\sqrt{2},-1/\sqrt{2})\,\sin{t}$. In the
case of rotation $B\neq 0$ and we expect asymptotically $\beta=1/4$, while
in the case of translations $B=0$ and we expect asymptotically
$\beta=1/6$. Numerical results shown in figure 3 confirm the
expectations very well. Additionally, we can see the transient regimes where
$\beta$ is different from asymptotic values. For small enough velocities we observe the regime
where $\beta=1/2$. This is because when the velocity is small the billiard
undergoes many oscillations between the collisions and the successive
velocity changes are effectively uncorrelated, leading to the random walk
like process. In the rotating case we see that after the random walk like
phase the system enters the intermediate regime where $\beta=1/6$, which
is because the asymptotically big term $B\,v'\varsigma$ is still much
smaller than $\dot{A}\,\eta$, regarding the equation (\ref{eqPowDef}). 
Eventually, for $n> 10^4$, $\beta$ becomes
$1/4$ as predicted by the theory.

Finally, we should add that numerical calculations have been performed for
various uniformly rotating billiards (Sinai billiard, elliptical billiard, 
Robnik billiard \cite{Robnik1983}, 
oval billiard \cite{Leonel2009a}) and the conservation law (\ref{eqConLawRotat}) has been
confirmed in double precision accuracy, which implies $\beta=0$.

\section{Conclusions}

In time-dependent fully chaotic shape-preserving billiards the velocity
dynamics is determined by the rotational properties of the billiard.
If the transformation is such that the angular momentum of the billiard is
preserved then the 
acceleration exponent is $\beta=1/6$, except in the case where the only
transformation is a uniform rotation and there is no acceleration, $\beta=0$. 
On the other hand, if the transformation is such that the
angular momentum of the billiard is not preserved then the acceleration
exponent equals $\beta=1/4$.
These three values of $\beta$ exhaust all possible values of the acceleration
exponent in fully chaotic time-dependent billiards. However, if the
structure of the phase space is more complicated, with the coexisting
islands of regular motion, the acceleration exponents may differ from the
prediction of the theory presented in this paper, simply because the
assumption for the autocorrelation functions
(\ref{eqBcorr1}), (\ref{eqBcorr2}) and (\ref{eqBcorr3}) are not fulfilled. 
In this work we address only the fully chaotic billiards, while
the results for the integrable ellipse and mixed type billiards in this
context will be published elsewhere.
The theory presented in this work complements the other
more general theories of the velocity dynamics in time-dependent
billiards \cite{Karlis2012,Gelfreich2008}. 

\section*{Acknowledgement}

I would like to thank Prof. Marko Robnik for a careful reading 
and great help in improving the quality of the paper.
This work was supported by the Slovenian Research Agency (ARRS).

\bibliography{fermiacc.bib}

\end{document}